\documentclass[twocolumn]{article}

\pdfoutput=1
\usepackage{jsaiac}
\usepackage{comment}
\usepackage[dvipdfmx]{color}
\usepackage[pdftex]{graphicx}
\usepackage{bm,enumerate,amsmath,amssymb,amsthm}
\usepackage{url}

\def \fext{pdf}
\def \vect#1{\mbox{\boldmath $#1$}}

\newcommand{\ms}{\mathcal}

\excludecomment{hdn}

\title{Coordination of Players in Ride-Sharing Games by Signaling}

\author{%
	Tatsuya Iwase\first{}
	\and
	Takahiro Shiga\first{}
}

\address{\first{}Toyota Central R\&D Labs., Inc.,	41-1 Yokomichi Nagakute, Aichi 480-1192, Japan \\ Email: tiwase@mosk.tytlabs.co.jp}

\affiliate{
	\first{}Social Systems Research-Domain \\
	Toyota Central R\&D Labs., Inc.
}

\begin{abstract}
Sharing unused vehicles is one practical solution for traffic congestion. We propose an advanced vehicle-sharing service that maximizes the sharing of vehicles and improves traffic efficiency by coordinating user trips via an information system. We formulate ride-sharing games that model externalities in vehicle sharing caused by insufficient vehicle supply. We show how Bayes correlated equilibrium can coordinate players in ride-sharing games and verify the resultant improvement in the price of anarchy.

\end{abstract}

\def\BibTeX{{\rm B\kern-.05em{\sc i\kern-.025em b}\kern-.08em%
		T\kern-.1667em\lower.7ex\hbox{E}\kern-.125emX}}
\def\JBibTeX{\leavevmode\lower .6ex\hbox{J}\kern-0.15em\BibTeX}
\def\LaTeXe{\LaTeX\kern.15em2$_{\textstyle\varepsilon}$}

\begin{document}
\maketitle
	
\begin{hdn}
	\color{red}
	\begin{verbatim}
	\end{verbatim}
	\color{black}

\end{hdn}
	
\section{Introduction}
\label{sec:intro}
	
Game theory seeks to solve the traditional problems of congestion and effective resource allocation. As populations increasingly concentrate in big cities, congestion becomes an increasingly critical problem. One practical solution for congestion is sharing unused or idled resources. For example, many vehicles inching along heavily congested roads have empty seats, or they occupy increasingly scarce urban parking spaces while pedestrians struggle to hail available taxis. There are many examples of fallow resources, including unoccupied buildings, empty restaurants, and unemployed workers.

We propose an advanced vehicle-sharing service that combines car sharing and ride sharing. People share their personal vehicles, and every user can be a driver or passenger of a shared vehicle instead of its owner. Their trips are coordinated by a mediator system to maximize the use of shared vehicles and improve transportation efficiency.

Considering the negative externalities caused by insufficient vehicle supply is the presiding difference between studies that examine vehicle sharing and those that examine traffic routing. While players drive their own vehicles in the traditional traffic-routing problem, players must locate vehicles to share before riding them in the vehicle-sharing problem. In the latter problem, players who change routes alter vehicle availability and impose externalities on other players---even those who are not transiting common routes. For example, consider a sequence in which Player 1 first drives a vehicle from A to B, next Player 2 drives that vehicle from B to C, and then Player 3 drives it from C to D. If Player 1 does not complete his/her route, Player 2 and Player 3 cannot use the vehicle even though Player 3 shares no part of his/her route with Player 1.

The congestion game introduced by Rosenthal\cite{rosenthal} has been applied to analyzing traffic congestion externalities\cite{cgapp}. Traffic engineers know that selfish route choice behavior diminishes the efficiency of transportation. Several studies employ congestion games to examine how traffic can be controlled by coordinating the behaviors of selfish drivers\cite{cgmech}. Others examine the loss in welfare created by selfish traffic. In other words, they compare the costs generated by selfish traffic to costs under optimally controlled traffic and refer to the difference as the {\it price of anarchy} (PoA)\cite{poa}. It seems that selfishness also reduces the efficiency of vehicle sharing. However, congestion games involve only externalities among players choosing partially identical routes; they are not useful in analyzing the externalities in vehicle sharing caused by inadequate vehicle supply. Agatz\cite{agatz} reviews studies of vehicle sharing and notes that they do not analyze PoA.

Hara et al.\cite{hara} apply mechanism design theory to the coordination of selfish users in vehicle sharing. However, the mechanism imposes complex computations of trip values and unlimited budgets on users, and it presents practical difficulties in implementation.

Signaling is another relatively new approach to coordination. A mediator provides players information to control their beliefs concerning uncertain environments when information asymmetry exists between the mediator and players\cite{kame,bce,kremer}. Accordingly, the mediator can control the expected payoff and the resulting choices of players. Signaling is easily implemented via a mobile phone app that provides information to users. Although several studies apply signaling to transportation problems\cite{rogers,vasserman}, they focus on traffic routing, not vehicle sharing.

This study involves coordinating selfish users of shared vehicles via signaling. It analyzes improvements in PoA in a manner similar to the analysis of traffic routing in congestion games.

\section{The Models}
\label{sec:model}

\subsection{Ride-sharing games}
\label{sec:rsg}

This section prepares an analytical tool for vehicle sharing similar to congestion games for traffic-routing analysis. We formulate ride-sharing games that model positive and negative externalities arising from vehicle supply. 

A {\it ride-sharing game} is defined as a tuple $G=<\ms{N},\ms{M},\ms{T},\ms{G},\ms{A},\mu,c>$. $\ms{N}=\{1,\ldots,N\}$ is a finite set of players. A player $i \in \ms{N}$ represents a user of shared vehicles. $-i$ represents all players except $i$. $\ms{M}=\{1,\ldots,M\}$ is a finite set of vehicles. Each vehicle $m \in \ms{M}$ has a common seating capacity $w \in \mathbb{N}_{> 0}$. 

$\ms{G}=<\ms{V},\ms{E}>$ is a directed graph featuring finite sets of nodes $\ms{V}=\{1,\ldots,V\}$ and edges $\ms{E}=\{1,\ldots,E\}$. $\ms{G}$ is a simple graph, but each node has a loop to itself. A node $v \in \ms{V}$ represents a place, and an edge $e \in \ms{E}$ represents a road. Players and vehicles move on $\ms{G}$.

$\ms{T}=\{1,\ldots,T\}$ is a finite set of time that partitions the day. Each player and vehicle is located on a node at time $t \in \ms{T}$ and finishes a move on an edge during period $(t,t+1)$.

$\ms{R}$ is a set of all paths with length $T-1$ on $\ms{G}$. A path $r=(v_{1},v_{2},\ldots,v_{T}) \in \ms{R}$ represents a player's roundtrip during a day. $\ms{A}_{i} \subset \ms{R}$ is a set of strategies of player $i$. $\ms{A}=\underset{i \in \ms{N}}{\times}\ms{A}_{i}$ is a set of strategy profiles. $a_{i} \in \ms{A}_{i}$ is a roundtrip of player $i$, and $\vect{a} \in \ms{A}$ is a strategy profile. $\vect{a}_{-i}$ represents a strategy profile of all players except $i$.

$\mu(i,t,\vect{a}):<\ms{N},\ms{T},\ms{A}> \to \ms{M}$ is a map that represents the allocation of player $i$ to vehicle $m$ during each period $(t,t+1)$ depending on strategy profile $\vect{a}$. If no vehicle is allocated to player $i$, $\mu(i,t,\vect{a})=\emptyset$. Each vehicle $m$ moves together with allocated player $i$ on the same edge where the player moves. $s_{m}(t,\vect{a})$ represents the number of players riding in vehicle $m$ during period $(t,t+1)$ when the strategy profile is $\vect{a}$.

$c_{e}(w,s_{m}):<\mathbb{N}_{> 0},\mathbb{N}_{\ge 0}> \to \mathbb{R}_{\ge 0}$ is a cost function of a player riding in vehicle $m$ on edge $e$. $c=\{c_{e}|e \in \ms{E}\}$ is a set of cost functions of all edges. The total cost of player $i$ in a day is $c_{i}(\vect{a})=\sum_{e_{t} \in a_{i}}c_{e}(w,s_{\mu(i,t,\vect{a})}(t,\vect{a}))$.

We consider one-shot games wherein players simultaneously choose entire roundtrips $\vect{a}$ during one day. We assume that the cost function $c_{e}$ is monotonically decreasing for $s_{m}$ when $s_{m} < w$ and monotonically increasing when $s_{m} \ge w$. We also assume $\mu$ so that users choose to ride vehicles selfishly to reduce their costs. An allocation mechanism is needed if the demand for vehicles exceeds supply on a node or edge, but that issue lies outside our scope of study.

\subsection{Bayesian ride-sharing games}
\label{sec:brsg}

Here we consider cases wherein players have incomplete information regarding vehicle allocation. A {\it Bayesian ride-sharing game} is an extension of a ride-sharing game and is defined as $G_{b}=<\ms{N},\ms{M},\ms{T},\ms{G},\ms{A},\ms{X},\ms{P},\mu,c>$.

$\ms{X}$ is a set of possible values of an exogenous variable $x \in \ms{X}$, which affects the allocation of vehicles $\mu$. $\mu(i,t,\vect{a}|x)$ is the allocation of vehicles depending on $x$. Similarly, $s_{m}(t,\vect{a}|x)$ is the number of players riding on vehicle $m$ depending on $x$, and $c_{i}(\vect{a}|x)$ is the cost to player $i$ depending on $x$. $p_{i}(x):\ms{X} \to [0,1]$ is a probability distribution of $X$ for player $i$, which represents his/her belief. $\ms{P}=\{p_{i}|i \in N\}$ is a set of probability distributions of all players. The definitions of other elements of $G_{b}$ are the same as those in ride-sharing game $G$.

Examples of the exogenous variable $x$ are initial vehicle locations and demand information aggregated through the Internet. Each player chooses $a_{i}$ to minimize his/her expected cost, which is $\mathbb{E}_{\ms{X}}[c_{i}]=\sum_{x \in \ms{X}}c_{i}(\vect{a}|x)p_{i}(x)$. If all players behave selfishly, the resulting strategy profile is a Bayesian Nash Equilibrium (BNE).

\subsection{Signaling in Bayesian ride-sharing games}
\label{sec:signal}

We use {\it Bayes Correlated Equilibrium} (BCE)\cite{bce} as a signaling technique used by a mediator to coordinate selfish players and improve the efficiency of vehicle sharing. BCE is a conditional distribution $\sigma(\hat{\vect{a}}|x)$ of a random recommendation $\hat{\vect{a}}$ that gives players an incentive to follow. Given the cost function $c_{s}(\vect{a}|x)$, the mediator's problem is to design an optimal recommendation that motivates players to coordinate to minimize the mediator's cost. The problem is expressed as follows:

\begin{equation}
	\left.
	\begin{array}{l}
		\max_{\sigma} \mathbb{E}_{x}[c_{s}(\hat{\vect{a}}|x)] \\
		s.t. \: \sum_{\hat{\vect{a}}_{-i},x}p_{i}(x)\sigma(\hat{\vect{a}}|x)c_{i}(\hat{a}_{i},\hat{\vect{a}}_{-i}|x) \leq \\
		 \sum_{\hat{\vect{a}}_{-i},x}p_{i}(x)\sigma(\hat{\vect{a}}|x)c_{i}(a_{i},\hat{\vect{a}}_{-i}|x),\: \forall i \forall \hat{a}_{i}.
	\end{array}
	\right.
	\label{eq:problem}
\end{equation}

The constraint represented above is called {\it incentive compatibility} (IC), and it renders every player unable to reduce his/her cost by deviating from the action recommended by the mediator.

\section{Examples}
\label{sec:exam}

\begin{figure}[h]
	\centering
	\includegraphics[clip,width=6cm]{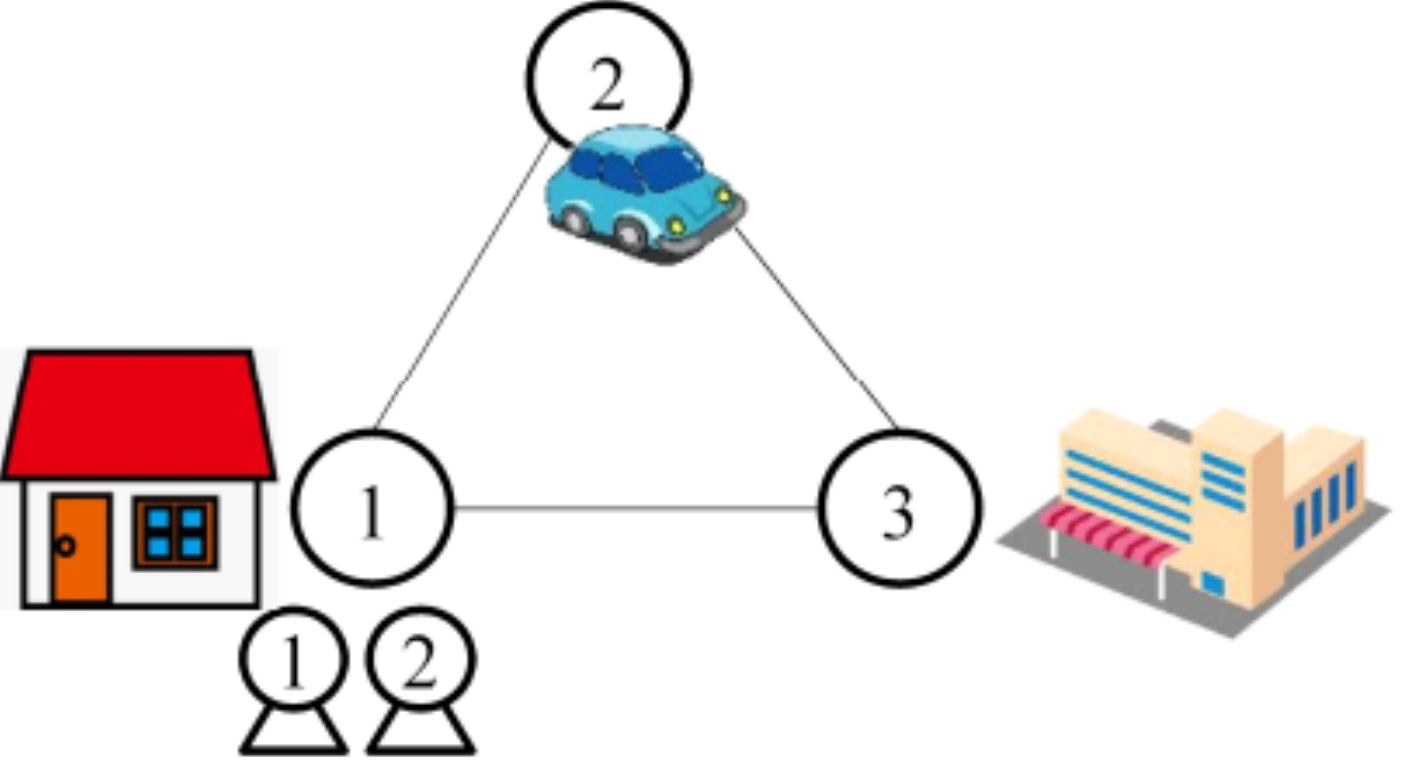}
	\caption{A Bayesian ride-sharing game}
	\label{fig:brsg}	
\end{figure}

Here, we show how signaling can improve the efficiency of sharing by incentivizing players to coordinate with each other in a Bayesian ride-sharing game. Game $G_{b}$ is defined as follows:

\begin{itemize}
	\item $N=2, V=3, T=4, M \le 1$.
	\item $\ms{G}$ and initial locations are shown in Figure \ref{fig:brsg}. All nodes have loop edges to themselves.
	\item $a_{i}$ must include node 3 for all players.
	\item A player uses the vehicle if it is located on his node.
	\item There is uncertainty $x \in \ms{X}=\{0,1\}$ regarding the availability of the vehicle. $x=0$ means $M=0$ and $x=1$ means $M=1$.
	\item All players have a common prior $p_{i}(x=0)=0.5, \forall i$. 
\end{itemize}

Each player has only two distinct options such that $\ms{A}=\{C,D\}$. $C=(1,2,3,1)$ is a trip that visits nodes in this order. On the other hand, $D=(1,1,3,1)$. All edges except for loop edges have the same cost function. If a player does not use the vehicle, the cost is 8. If a player drives alone, the cost is 6. If two players share the vehicle, the cost is 1. The cost at loop edges is 0. Cost matrices of this game appear in Tables \ref{tbl:cost0} and \ref{tbl:cost1}.

The expected cost matrix is shown in Table \ref{tbl:ecost}. It has the structure of a prisoner's dilemma because each player prefers to go to node 3 without picking up the car and return home riding with another player rather than risk picking up the car in person. Accordingly, BNE is $a=(D,D)$, which means that no players share the vehicle.

Now we consider a system that uses BCE to coordinate players to share the unused vehicle. The system cost can be denoted as $c_{s}(a|x)=\sum_{i \in \ms{N}}c_{i}(a|x)$. The problem of the system is denoted by Eq.\ref{eq:problem}, which is the search for an optimal recommendation policy $\sigma(\hat{\vect{a}}|x)$ as in Table \ref{tbl:sigma}. The problem becomes one of linear programming, and Table \ref{tbl:sigmaopt} presents a solution. This incentive-compatible recommendation induces the coordination of players and realizes BCE, where the system's expected cost $\mathbb{E}_{x}[c_{s}(\hat{\vect{a}},x)]=27.9$, which is better than the one under BNE, $\mathbb{E}_{x}[c_{s}(\vect{a}|x)]=32$. Since $\mathbb{E}_{x}[c_{s}(\vect{a}|x)]=26$ in social optimum, the PoA is improved from 1.23 under BNE to 1.07 under BCE.

\begin{table}[!]
	\centering
	\caption{$c_{i}(a_{1},a_{2}|x=0)$}
	\begin{tabular}{|c|c|c|} \hline
		& C & D \\ \hline \hline
		C & 20,20 & 20,16 \\ \hline
		D & 16,20 & 16,16 \\ \hline
	\end{tabular}
	\label{tbl:cost0}
	
	\caption{$c_{i}(a_{1},a_{2}|x=1)$}
	\begin{tabular}{|c|c|c|} \hline
		& C & D \\ \hline \hline
		C & 10,10 & 15,9 \\ \hline
		D & 9,15 & 16,16 \\ \hline
	\end{tabular}
	\label{tbl:cost1}
\end{table}

\begin{table}[!]
	\centering
	\caption{$\mathbb{E}_{x}[c_{i}(a_{1},a_{2}|x)]$}
	\begin{tabular}{|c|c|c|} \hline
		& C & D \\ \hline \hline
		C & 15,15 & 17.5,12.5 \\ \hline
		D & 12.5,17.5 & 16,16 \\ \hline
	\end{tabular}
	\label{tbl:ecost}
\end{table}


\begin{table}[!]
	\centering
	\caption{$\sigma(\hat{a}_{1},\hat{a}_{2}|x)$}
	\begin{tabular}{|c|c|c||c|c|c|} \hline
		\multicolumn{3}{|c||}{$\sigma(\hat{a}_{1},\hat{a}_{2}|0)$}  & \multicolumn{3}{|c|}{$\sigma(\hat{a}_{1},\hat{a}_{2}|1)$} \\ \hline \hline
		& C & D &  & C & D \\ \hline
		C & $\alpha_{0}$ & $\beta_{0}$ & C & $\alpha_{1}$ & $\beta_{1}$ \\ \hline
		D & $\beta_{0}$ & $1-\alpha_{0}-2\beta_{0}$ & D & $\beta_{1}$ & $1-\alpha_{1}-2\beta_{1}$ \\ \hline
	\end{tabular}
	\label{tbl:sigma}
\end{table}

\begin{table}[!]
	\centering
	\caption{Optimal $\sigma(\hat{a}_{1},\hat{a}_{2}|x)$}
	\begin{tabular}{|c|c|c||c|c|c|} \hline
		\multicolumn{3}{|c||}{$\sigma(\hat{a}_{1},\hat{a}_{2}|0)$}  & \multicolumn{3}{|c|}{$\sigma(\hat{a}_{1},\hat{a}_{2}|1)$} \\ \hline \hline
		& C & D &  & C & D \\ \hline
		C & 0 & 0 & C & 0.06 & 0.47 \\ \hline
		D & 0 & 1 & D & 0.47 & 0 \\ \hline
	\end{tabular}
	\label{tbl:sigmaopt}
\end{table}

\section{Conclusion}
\label{sec:conc}

\begin{hdn}
	\color{red}
	\begin{verbatim}
	\end{verbatim}
	\color{black}
\end{hdn}

We have formulated ride-sharing games that model externalities caused by insufficient vehicle supply. We verified the coordination of selfish users by BCE and improvements in PoA on the basis of a simple ride-sharing game. Future studies should determine the theoretical bounds of PoA and propose an approximate algorithm for general games followed by practical verification.

\end{document}